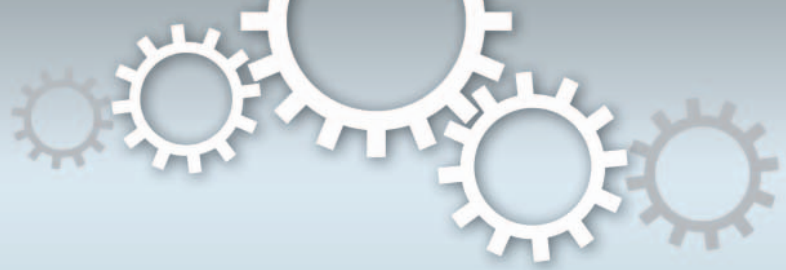

OPEN

# Do cancer cells undergo phenotypic switching? The case for imperfect cancer stem cell markers

Stefano Zapperi[1,2] & Caterina A. M. La Porta[3]

[1]CNR-IENI, Via R. Cozzi 53, 20125 Milano, Italy, [2]ISI Foundation, Via Alassio 11/C, 10126 Torino, Italy, [3]Department of Life Sciences, University of Milano, via Celoria 26, 20133 Milano, Italy.

The identification of cancer stem cells in vivo and in vitro relies on specific surface markers that should allow to sort cancer cells in phenotypically distinct subpopulations. Experiments report that sorted cancer cell populations after some time tend to express again all the original markers, leading to the hypothesis of phenotypic switching, according to which cancer cells can transform stochastically into cancer stem cells. Here we explore an alternative explanation based on the hypothesis that markers are not perfect and are thus unable to identify all cancer stem cells. Our analysis is based on a mathematical model for cancer cell proliferation that takes into account phenotypic switching, imperfect markers and error in the sorting process. Our conclusion is that the observation of reversible expression of surface markers after sorting does not provide sufficient evidence in support of phenotypic switching.

E vidence indicating that tumors are composed by a heterogeneous cell population has accumulated for long time[1]. There are two different general hypothesis on the nature of this heterogeneity: the first states that cancer cells might differ but all cells are potentially tumorigenic (conventional model) while the second states that only a subset of cells, the cancer stem cells (CSCs), are tumorigenic and drive tumor growth (hierarchical model)[2]. CSCs are usually identified using serial transplantation, validating a candidate CSC subpopulation by monitoring the capability to recapitulate the heterogeneity of the primary tumor. Both xeno- and syngeneic transplantation might, however, misrepresent the real intricate network of interactions with diverse supports such as fibroblasts, endothelial cells, macrophages, mesenchymal stem cells and many of the cytokines and receptors involved in these interactions (for a more comprehensive discussion read Ref.[3]). In addition, the success of this strategy is linked to the choice of an appropriate marker that can correctly identify the CSC population both in xenografts and in biotic samples. Due to these problems, the presence of CSC in solid tumors is still debated.

In this context a recently proposed hypothesis states that phenotypes in a cancer cell population are not static but can switch stochastically[4]. The idea underlying this phenotypic switching hypothesis is that any biological system is subject to a varying degree of noise in key signalling pathways that may lead to heritable changes in gene expression through epigenetic mechanisms[5,6]. To prevent that this noise could trigger an inappropriate cellular response, signalling systems may be buffered in such a way that the cells would respond to yield a specific biological output, such as a switching its phenotype, only when a critical signalling threshold is crossed. In cancer cells a phenotype instability could be due to genetic lesions that constitutively activate one signalling pathway playing a key role in buffering the output from a second pathway leading the cells to become more sensitive to microenvironment. According to this idea, phenotypic switching in cancer cells may reflect a lowering of the threshold necessary to trigger a change in cell identity in response to external signals originating within the tumor microenvironment that may vary substantially from location to location. Hence, if phenotypic switching is reversible, most cells should have the potential to adopt a stem cell like phenotype accounting for the high proportion of cells able to seed tumors in severely immunocompromise animals[7,8]. In a recent paper Gupta et al. show that subpopulations of breast cancer cells of a given phenotypic state over time express again all the original phenotypes. These results are interpreted by a simple Markov model involving a tiny probability for cancer cells to switch back to the CSC state[4]. Other papers, however, do not support the phenotypic switching hypothesis. In melanoma, ABCB5 cells are not able to generate ABCB5⁺ cells[9], CD34⁺Cd271/Ngfr/p75 cells formed tumors CD271 restricted, whereas CD34CD271/Ngfr/p75 cells formed tumors containing both CD271⁺ and CD271 cells[10].

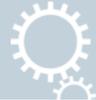

From the biological point of view, it is not easy to determine if the tumor grows following the conventional or the hierarchical model and to understand the nature of phenotypic switching. In this respect, mathematical models can prove very useful to clarify the consequences of biologically motivated assumptions. The key issue is to explain how a purified subpopulation can express CSC markers after sorting. A possible explanation is provided by the phenotypic switching hypothesis: if phenotypes evolve dynamically it is possible that cells originally negative to the CSC phenotype may express it later due to stochastic fluctuations (See Fig. 1A). This explanation is somewhat problematic from the conceptual point of view: if cancer cells (CCs) can transform back into CSCs then the very notion of CSC becomes blurred. A key distinction between CSCs and other CCs is that the first generate the latter and not vice versa. Furthermore, CSC should be virtually immortal while CCs should stop replicating after a finite number of divisions. Once we accept that CCs can return to the CSC state, they become potentially immortal as well. Hence the distinction between phenotypic switching and the original conventional model risks of becoming purely semantic.

In this paper, we explore a conceptually simple alternative to explain the experimental results. The operative identification of CSCs relies on stem cell markers, but their absolute accuracy is far from being guaranteed. Our proposal is to assume that putative CSC markers are imperfect and derive the experimental consequences of this assumption (See Fig. 1B). An imperfect marker yields a marker-positive subpopulation that is CSC rich, but does not allow to eliminate all CSCs from the marker-negative subpopulation. The few CSCs present in the marker-negative subpopulation will drive tumor growth and re-establish a marker-positive subpopulation in the tumor. This scenario was ruled out in Ref. [4] because the short-term growth rate for stem-like and non-stem-like cells was simular. Here we demonstrate that this observation can also be explained by the imperfect marker hypothesis, since a difference in growth is only observable in the long-term, as also shown in Ref. [11]. Our conclusion is that current experimental data that have been claimed to support the phenotypic switching hypothesis[4] can also be interpreted within an imperfect marker scenario.

To model the kinetics of cancer cells, we employ a standard population dynamics approach, using the theory of branching processes[12,13]. Branching processes have been used extensively in the last decades to model the growth of stem cells[14–19] and more recently of cancer stem cells[20–23]. In this paper, we employ and extend the branching process model for CSCs discussed in Ref. [11]. The main limitation of branching processes is due to their mean-field nature that does not take into account the geometry of the cellular arrangement inside a tumor. Despite this shortcoming, the model allows for a quantitative description of experimental growth curves of melanoma cells, providing an indirect confirmation of the CSC hypothesis in this tumor[11]. It would be interesting to understand better CSC localization, in analogy with stem cells in tissues where it is possible to study their spatio-temporal kinetics in vivo[24,25], using statistical mechanics models to understand the results[26]. Similar techniques for cancer are still at a preliminary stage and have been use to track metastatic cells in vivo[27]. Tracking CSCs appears to be more complicated mainly due to the lack of unambiguous markers.

## Results

**Simple Markov model.** The experimental results on breast cancer cells reported in Ref. [4] were originally interpreted by a simple Markov model. Cancer cells were sorted in three classes (Stem-like, basal and luminal) depending on the expression of a combination three markers (CD24,CD44 and EpCAM). By defining relative fractions $f_i(t)$ for each class at time $t$, the time evolution of each population was chosen to follow the equation

$$f_i(t) = \sum_j P_{ij}f_j(t-1), \qquad (1)$$

where time is measured in days, and $P_{ij}$ is the probability per unit day that a cell of type $j$ transforms into a cell of type $i$. This equation has a formal solution

$$f_i(t) = \sum_j (P^t)_{ij}f_j(0), \qquad (2)$$

where $P^t$ is the $t$ – power of the matrix $P$. In Ref. [4], an estimate for the matrix $P$ was obtained by sorting the cells into three classes and then sorting each subpopulation again after six days of cultivation.

Here, we consider the simple case in which cells are sorted in only two classes: CSCs and CC. Taking advantage of the normalization condition $\Sigma_i \rho_i = 1$, the evolution equation in Eq. 1 can then be written in terms of the density of CSC $f_1$ alone

$$f_1(t) = P_{11} f_1(t-1) + P_{21}(1 - f_1(t-1)) \quad (3)$$

where $P_{11}$ is the probability per day that a CSC remains a CSC and $P_{21}$ is the probability that a CC transforms to a CSC. Eq. 3 has explicit solution

$$f_1(t) = f_1(0)(P_{11} - P_{21})^t + P_{21} \frac{1 - (P_{11} - P_{21})^t}{1 - P_{11} + P_{21}}. \quad (4)$$

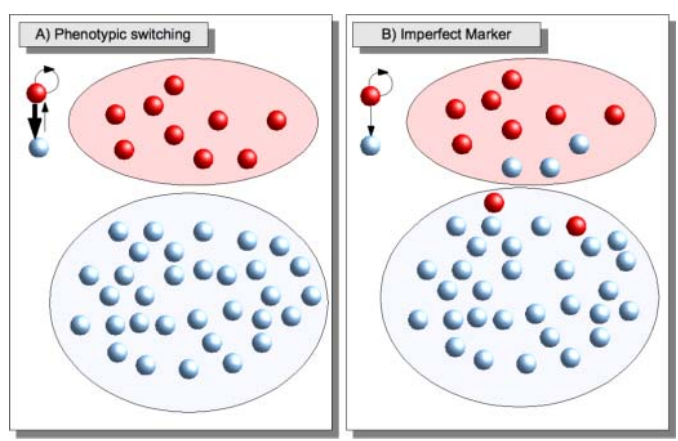

**Figure 1 | Phenotypic switching and Imperfect markers.** (A) According to the phenotypic switching hypothesis, CCs (blue) have a small probability to revert to the CSC state (red). If a marker is used to sort the cells into different subpopulation, the negative subpopulation will eventually express again the marker due to phenotypic switching. (B) According to the imperfect marker idea, CCs can not transform back into CSCs, but both CCs and CSCs express the marker, although in different proportions: most of the CSCs are positive, while most of the CCs are negative.

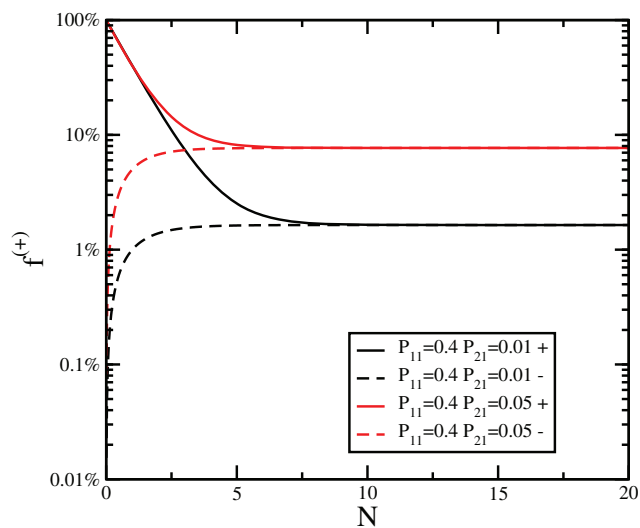

**Figure 2 | Evolution of the concentration of positive cells after sorting in the Markov model.** The evolution of the concentration of positive cells after sorting for positive (+) and negative (−) subpopulations as a function of the number of generations $N$ for the Markov model with $P_{11} = 0.4$ and two different values of $P_{21}$.

At long times the fraction of CSCs is given by $f_1^\infty = P_{21}/(1 - P_{11} + P_{21})$ and the steady-state is reached exponentially with a typical timescale $\tau = -1/\log(P_{11} - P_{21})$ both for positive ($f_1(0) = 1$) and negative sorted subpopulations ($f_1(0) = 0$). An illustration of the behavior of the model is reported in Fig. 2.

The model is particularly simple but does not really distinguish between CSCs and CCs, since all cell classes are treated in the same way and proliferation is not accounted for. Therefore, the simple Markov model describes cells that are heterogeneous but not hierarchically organized as in the conventional cancer model. It is, however, possible to combine in a mathematical model phenotypic switching with a hierarchical organization of the cells as we will discuss below.

**CSC model.** We consider a stochastic model for the proliferation of hierarchically organized cancer cells introduced in Ref. [11] and illustrated in Fig. 3A. According to the CSC hypothesis, cells are organized hierarchically, with CSCs at the top of the structure. CSCs can divide symmetrically giving rise to two new CSCs with probability $\epsilon$ or asymmetrically with probability $1 - \epsilon$ giving rise to a CSC and a CC. While CSCs can duplicate for an indefinite amount of time, CCs become senescent and stop duplicating after a finite number of generations $M$. This is the minimal ingredient needed to model the CSC hierarchy. It is possible that CSCs differ in other biological aspects from CCs, but this is irrelevant from the point of view of population dynamics. This model was successfully used to describe the growth of melanoma cells, where the best fit to the data yields $\epsilon \simeq 0.7$ and $M = 38$[11].

The analytical solution of the CSC model has been reported in Ref. [11], writing down the equations for the evolution of cell populations

that have a recursive form[11]

$$S^N = (1 + \epsilon)S^{N-1}$$
$$C_1^N = (1 - \epsilon)S^{N-1}$$
$$\dots \quad \dots \quad \dots \quad (5)$$
$$C_k^N = 2C_{k-1}^{N-1}$$
$$D^N = D^{N-1} + 2C_M^{N-1}.$$

Here $S^N$ is the number of CSCs after $N$ generations, $C_k^N$ is the number of CCs of "age" $k$ (i.e. that have undergone already k divisions) at generation $N$ and $D^N$ is the number of senescent cells at generation $N$. Solving Eq. 5 one can show that the asymptotic fraction of CSC is given by

$$f_{CSC} = \epsilon \left( \frac{1 + \epsilon}{2} \right)^M. \quad (6)$$

The time evolution can be obtained by introducing the division rate $R_d$, that for simplicity we set to be the same for CSCs and CCs. Time is then related to generation number by the relation $N = t R_d$. Eqs. 5 can be solved exactly to yield the number of cells in each class in terms of the initial conditions[11], or alternatively they can be evaluated numerically. In the CSC model CCs never revert to the CSC state and therefore a perfectly sorted CC population should always remain negative. On the other hand, if some CSCs are present the CSC fraction will eventually return to the steadystate value. Finally, the total number of cells is proportional to the number of CSC growing in time as

$$N(t) \propto (1 + \epsilon)^{R_d t}, \quad (7)$$

where $R_d$ is the rate of cell division per day.

**Phenotypic Switching model.** To introduce phenotypic switching into the CSC model, we consider the possibility for CCs to revert back the CSC state. This is done by introducing the probability $p$ that, instead of dividing, a CC transforms into a CSC (see Fig. 3B). Hence, CCs switch with probability $1 - p$ giving rise to two CCs as in the CSC model. This model of phenotypic switching retains the distinction between CSCs and CCs, since only the latter turn senescent after a

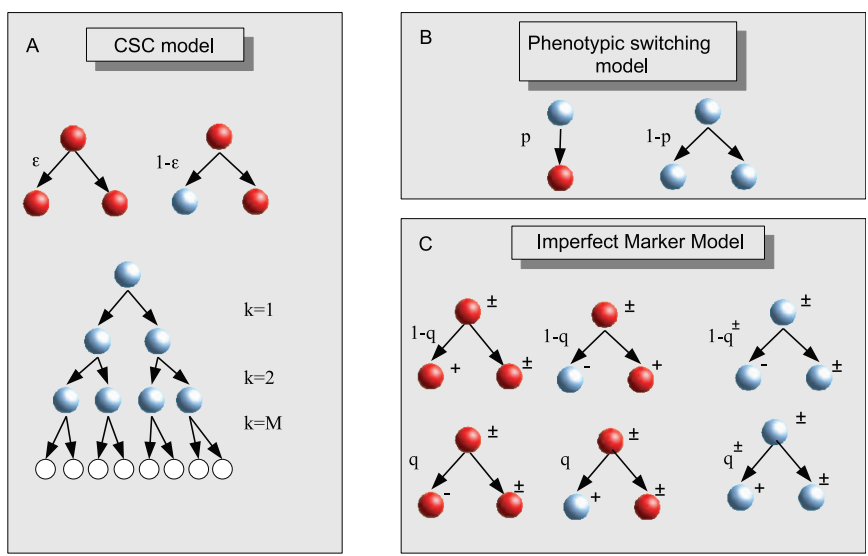

**Figure 3 | Models.** (A) In the CSC model, CSCs (red) can divide symmetrically yielding two CSCs with probability $\epsilon$ or asymmetrically yielding a CSC and a CC with probability $1 - \epsilon$. CCs divide symmetrically for $M$ generation after which they turn senescent. (B) Phenotipic switching is modeled by introducing a probability $p$ that a CC transform back to the CSC state instead of duplicating. (C) In the imperfect marker model, the switching concerns marker expression not the CSC state. Both CSCs and CCs can be positive to the marker and upon division the expression of the marker can change randomly with respect to the originating cell according to the probabilities $q$ for CSCs and $q^\pm$ for positive and negative CCs.

 

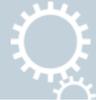

fixed number of divisions, unless they transform back to the CSC state. Finally, senescent cells are not allowed to switch to the CSC state.

To solve the model, we have to modify the recursion relations Eq. 5 to take into account the possibility that CCs revert to the CSC state with probability $p$. This leads to a set of equations

$$
\begin{aligned}
S^N &= (1+\epsilon)S^{N-1} + p\sum_{i=1}^{M} C_i^{N-1} \\
C_1^N &= (1-\epsilon)S^{N-1} \\
&\ldots \;\; .. \;\; \ldots \\
C_k^N &= 2(1-p)C_{k-1}^{N-1} \\
D^N &= D^{N-1} + 2(1-p)C_M^{N-1},
\end{aligned}
\tag{8}
$$

that we integrate numerically for different initial conditions, corresponding to positive and negative sorting. Typically we first determine the steadystate distribution $(S^\infty, C_k^\infty, D^\infty)$ and then perform the sorting by choosing negative cells as $(S^0 = 0, C_k^0 = C_k^\infty, D^0 = D^\infty)$ and positive cells as $(S^0 = S^\infty, C_k^0 = 0, D^0 = 0)$. We then evolve the system until it reaches the steadystate again. Fig. 4 illustrates the behavior of the model by following the fraction of positive cells $f^+$ in the two subpopulation for different values of $p$.

**Imperfect Marker model.** In the imperfect marker model, we assume that the marker does not allow to sort all the CSCs but can at most separate the cells into CSC rich and CSC poor populations. We start again from the CSC model and introduce a set of probabilities defining the evolution of the marker for CSCs and CCs (see Fig. 3C). At each cell division, CSCs have a probability $q$ of giving rise to one negative CSC or to one positive CC, while with probability $1 - q$ they give rise to a positive CSC or a negative CC. The other cell in the division process always retains the marker of the originating cell. If $q$ is small, then most CSCs will be positive and most CCs will be negative, while for $q = 0$ the marker is perfect. Similar rules apply for CCs: positive CCs have a probability $q^+$ to generate a positive CC and a probability $1 - q^+$ to generate a negative CC upon division, while the other cell remains positive. Negative CCs have yield a positive CC with probability $q^-$ and a negative CC with probability $1 - q^-$, while the other cell remains negative. While the results will crucially depend on the choice made for $q$, $q^+$ and $q^-$, here we consider only two extreme cases:

i) CSCs and CCs can change the expression of the marker in exactly the same way. This case corresponds to $q^+ = q^- = q$.

ii) Only CSCs can divide into cells that have a different expression of the marker with respect to the generating cells (with probability $q$), while CCs retain their marker upon division. This case corresponds to $q^+ = 1$ and $q^- = 0$: positive CCs always divide into positive CCs, while negative CCs always yield negative CCs.

The advantage of these two cases is that they involve a single parameter, $q$.

For the imperfect maker model, we have to write two sets of recursion relations for positive and negative cells:

$$
\begin{aligned}
S^{N(+)} &= (1+\epsilon(1-q))\left(S^{N-1(+)} + S^{N-1(-)}\right) \\
C_1^{N(+)} &= (1-\epsilon)q\left(S^{N-1(+)} + S^{N-1(-)}\right) \\
&\ldots \ldots \ldots \\
C_k^{N(+)} &= (1+q^+)C_{k-1}^{N-1(+)} + q^- C_{k-1}^{N-1(-)} \\
D^{N(+)} &= D^{N-1(+)} + (1+q^+)C_{k-1}^{N-1(+)} + q^- C_{k-1}^{N-1(-)},
\end{aligned}
\tag{9}
$$

and

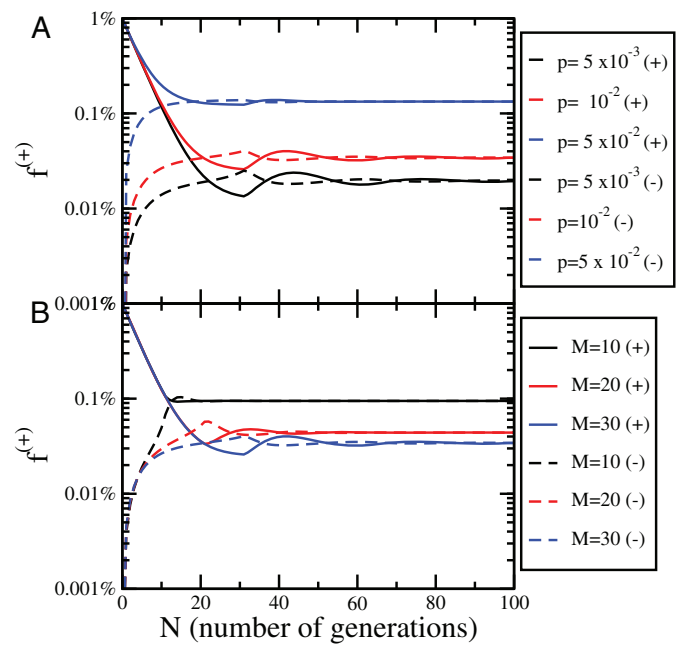

**Figure 4 | Evolution of the concentration of positive cells after sorting for the Phenotypic Switching model.** (A) The evolution of the concentration of positive cells after sorting for positive ($+$) and negative ($-$) subpopulations as a function of the number of generations $N$ for different values of the parameter $p$, $M = 30$ and $\epsilon = 0.6$. (B) The same plot as panel (A) for $p = 10^{-2}$ and different values of $M$.

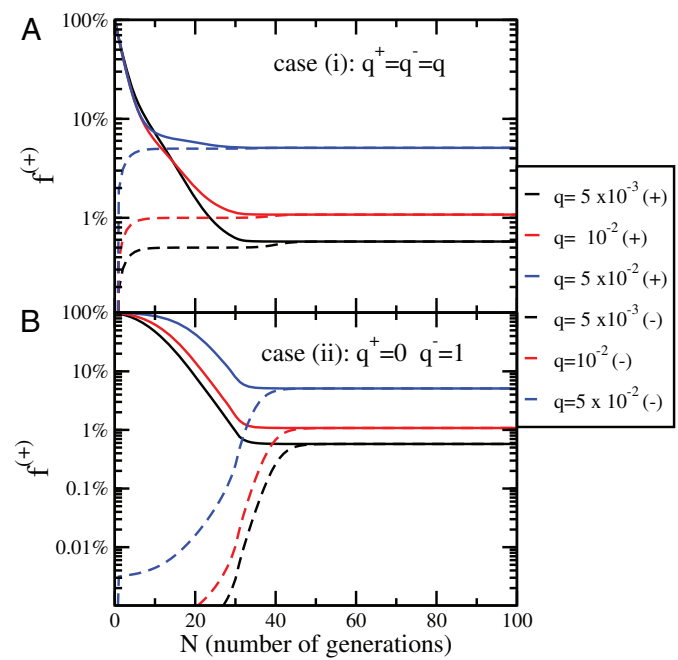

**Figure 5 | Evolution of the concentration of positive cells after sorting for the Imperfect Marker model.** (A) The evolution of the concentration of positive cells after sorting for positive ($+$) and negative ($-$) subpopulations as a function of the number of generations $N$ in case (i) ($q^+ = q^- = q$) for different values of the parameter $q$, $M = 30$ and $\epsilon = 0.6$. (B) The same plot as panel A) for case (ii) ($q^+ = 0$, $q^- = 1$) and different values of $q$.

$$S^{N(-)} = \epsilon q \left( S^{N-1(+)} + S^{N-1(-)} \right)$$
$$C_1^{N(-)} = (1-\epsilon)(1-q)\left( S^{N-1(+)} + S^{N-1(-)} \right)$$
$$\cdots\cdots$$
$$C_k^{N(-)} = (1-q^+)C_{k-1}^{N-1(+)} + (2-q^-)C_{k-1}^{N-1(-)} \qquad (10)$$
$$D^{N(-)} = D^{N-1(-)} + (1-q^+)C_{k-1}^{N-1(+)} + (2-q^-)C_{k-1}^{N-1(-)}.$$

In Fig. 5, we report the evolution of the fraction of positive cells in the sorted population for case (i) and (ii) as a function of $q$. In both the fraction of positive cells converges to the steady-state value with a timescale set by $M$ while the steady-state value depends on $q$.

**Dynamics after an imperfect sorting.** The last case considered is that of a perfect marker for CSCs, but an imperfect sorting. Sorting by Fluorescence-activated cell sorting (FACS) typically involves errors: some cells could be assigned to the wrong category. We measure the efficiency of the sorting by $\eta$, the probabilty that a cell is sorted incorrectly by FACS. An imperfect sorting on a cell population characterized by a number $[S^0, C_k^0, D^0]$ of CSCs, CCs and senescent cells yields a positive subpopulation composed by $[(1-\eta)S^0, \eta C_k^0, \eta D^0)]$ cells and a negative subpopulation composed by $[\eta S^0, (1-\eta)C_k^0, (1-\eta)D^0)]$ cells. Consequently the fraction of positive cells in the original population is not equal to the fraction of CSCs but is given by

$$f^{(+)} = (1-\eta)f_{CSC} + \eta(1-f_{CSC}), \qquad (11)$$

and only for $\eta = 0$ we have $f^{(+)} = f_{CSC}$. Using this model we can study the evolution of positive cells in sorted subpopulations.

To quantify the effect of an imperfect sorting, we consider the evolution of the concentration of positive cells as a function of the sorting efficiency $\eta$. Using the CSC model, we start from steady-state concentrations of CSCs and CCs and sort them into two subpopulations according to Eq. 11. Next, we integrate Eqs. 5 and at each generation we compute the fraction of positive cells. The result also in this case is that after some time the system returns to the steady state. As illustrated in Fig. 6A for $M = 30$ and $\epsilon = 0.8$, the evolution depends on $\eta$ only for the negative subpopulation and is independent on $\eta$ for the positive subpopulation. In both cases, the number of generations needed to reach the steady state is controlled by $M$, as shown in Fig. 6B. Hence, we can estimate the typical equilibration time to be around $t^* \simeq MR_d$ for the positive subpopulation and slightly larger for the negative one. The main difference between imperfect sorting and imperfect marker or phenotipic switching is that in the first case there is a net asymmetry between positive and negative subpopulations: the negative subpopulation remains roughly constant for the first $M$ generations, while the positive subpopulation decreases from the beginning.

The sorting efficiency, while not directly accessible from experiments, can be estimated by a simple calculation. When discussing FACS experiments it is customary to report the purity of the process, obtained by sorting the subpopulation immediately after the first sorting. Here we define the purity $\kappa$ of the sorting as the real concentration of positive cells present in the nominally positive subpopulation and express it in terms of $\eta$

$$\kappa = \frac{(1-\eta)f_{CSC}}{(1-\eta)f_{CSC} + \eta(1-f_{CSC})} \qquad (12)$$

Combining Eq. 11 and Eq. 12, we can estimate the sorting efficiency in terms of the measured values of $f^{(+)}$ and $\kappa$. In the limit $f^{(+)} \ll 1$, typical of CSC markers, and high purity $(1-\kappa \ll 1)$, we obtain a simple expression

$$\eta \simeq f^{(+)}(1-\kappa). \qquad (13)$$

To make a concrete example, Ref. [4] reports a purity of 96% and 2% stemlike cells obtained from SU159 breast cancer cells. Inserting $\kappa = 0.96$ and $f^{(+)} = 0.02$ in Eq. 13, we estimate $\eta = 8 \times 10^{-4}$.

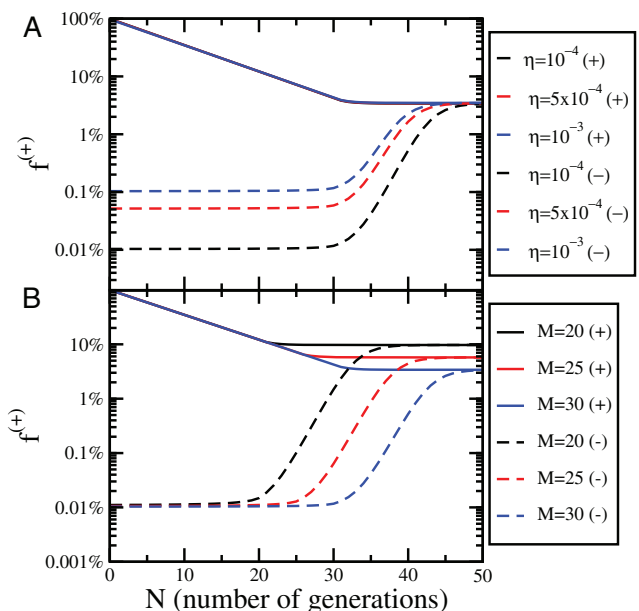

**Figure 6 | Evolution of the concentration of positive cells after an imperfect sorting.** (A) The evolution of the concentration of positive cells after sorting for positive ($+$) and negative ($-$) subpopulations as a function of the number of generations $N$ for different values of the sorting efficiency $\eta$. The dynamics is obtained solving the CSC model with $M = 30$ and $\epsilon = 0.8$. (B) The same plot as panel (A) for $\eta = 10^{-4}$ and different values of $M$.

**Comparison with experiments.** As discussed above, the simple Markov model, the CSC model with phenotypic switching, imperfect markers or imperfect sorting all yield the same outcome: after some time the fraction of cells that are positive to the marker returns to the original value. This proves that the expression of a putative CSC marker after positive cells have been eliminated by sorting is not a sufficient proof of phenotypic switching. To illustrate this point more clearly, we consider the experimental results reported by Gupta et al[4] on breast cancer cell lines. In Fig. 7 we report the fraction of positive (stem-like) cells six days after the initial sorting. These data were interpreted in Ref. [4] by the simple Markov model. Here we show that the same data can be reproduced by the imperfect marker model or by the phenotypic switching model. To this end we have chosen parameters so that the asymptotic value of $f^+$ is equal to the initial value, $f^+ = 1.9\%$, obtained in Ref. [4], thus assuming that the cell populations were originally in the steady-state. We notice that the experimental data could be interpreted as a result of an imperfect sorting if we assume that $M \sim 5$, which appears to be too small.

In order to get more insight on the process underlying the observed behavior one can consider the growth curves of the sorted subpopulations. The experimental data reported in Ref. [4] refer to two days of growth and show no significant difference in proliferation between sorted subpopulations. This observation was considered in Ref. [4] as additional evidence in favor of phenotypic switching. This is, however, not the case as shown in Fig. 8 which compares the experimental data with the prediction of the phenotypic switching and imperfect marker models. Both models show no difference in growth at short times, while the difference can only be observed at later times. A similar result was reported for melanoma cells sorted with a putative stem cell marker (ABCG2)[11]: a difference in the growth for positive and negative cells was observed only after two months of cultivation.

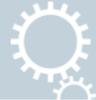

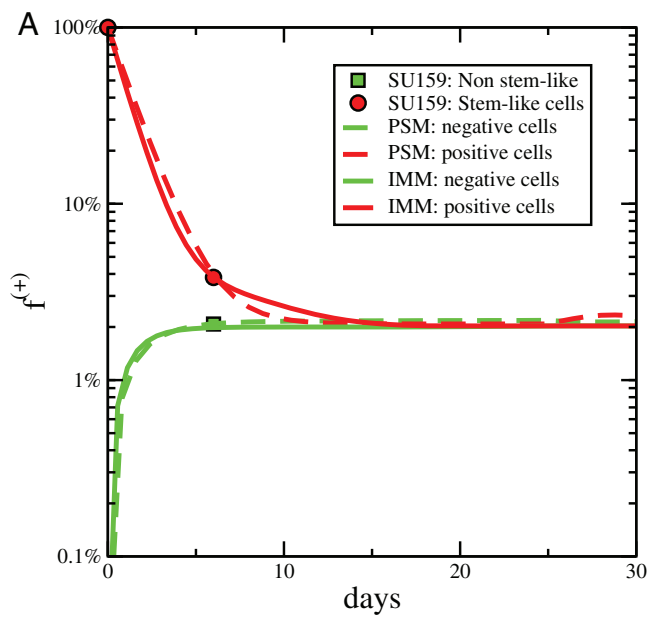

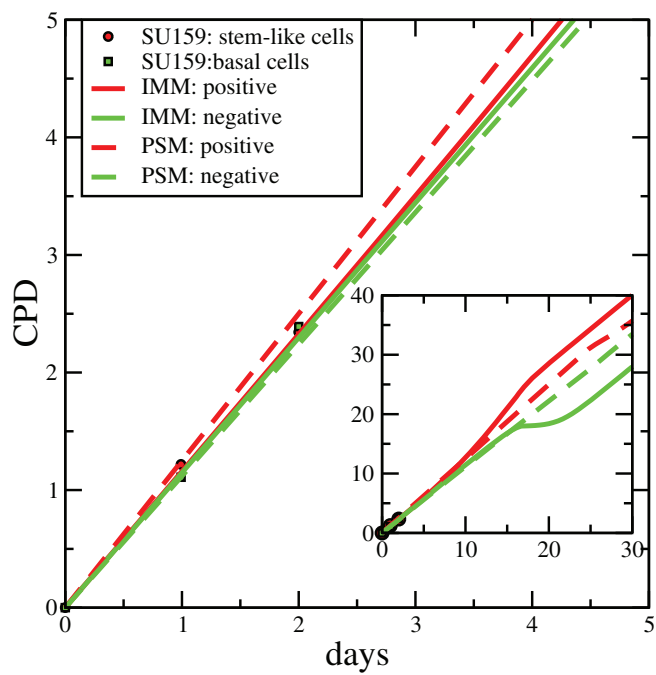

**Figure 7 | Comparison with experiments.** The evolution of the fraction of positive cells after sorting for positive (+) and negative (−) subpopulations as a function of time. Experimental data are extracted from Ref. [4] and compared with results of the phenotipic switching model (with $M = 30$, $\epsilon = 0.2$, $p = 0.0085$ and $R_d = 1.8$) and with the imperfect marker model (case (ii) with $M = 30$, $\varepsilon = 0.55$, $q = 0.02$ and $R_d = 1.8$).

**Figure 8 | Growth curves.** Using the same parameters employed in Fig. 7 one can also reproduce with phenotypic switching and imperfect marker models the growth curves of the two subpopulations. The differences between the two models appear only at long times (inset).

## Discussion

In the last decades, the main biologically motivated strategy to kill tumor cells has been to target common factors involved in cellular proliferation. The underlying idea was that if all cells are able to give rise to a tumor, one should try to identify the best factors that might affect the biological function of all the cells in order to kill them all together. Since tumor cells can proliferate indefinitely, the best candidates were supposed to be key factors involved in cellular division. Many factors have been claimed in the past to affect tumor proliferation, but most of time their clinical impact were modest in comparison to the effects demonstrated in vitro. Thus, either the factors were not the right ones or one should reconsider the traditional view of cancer.

In 1997, Bonnet et al[2] proposed that only a subpopulation of the cells can sustain tumor growth. These cells were defined CSCs because they share with stem cells biological characteristics like the unlimited capability to grow[3]. This observation changes completely the therapeutic perspective since the only way to eradicate the tumor is to target CSCs[28]. Yet, the identification of the best markers to define CSCs is an extremely controversial topic in the literature. For instance, Gupta et al[4] used CD44$^{high}$/CD24$^{−}$/EpCAM$^{low}$ to identify breast cancer CSC. In a different paper, however, CD44$^{+}$/CD24$^{low}$ cells were shown to be more abundant in triple-negative invasive breast carcinoma phenotype and to be associated with poor outcome[29]. In melanoma the story is quite similar with different markers used to identify CSCs[7–9,30–36] and no consensus on which one is the most effective. In a recent paper, Roesch et al used JARID1B to identify CSC and suggested that melanoma cells are not hierarchically distributed since the JARID1B$^{−}$ subpopulation can become positive after some time[8], as also observed by Quintana et al.[7] using other CSC markers and by Gupta et al[4] in breast cancer. These results lead to a new hypothesis that is gaining traction in the literature[37]: the possibility that CCs can switch back to the CSC phenotype.

In this paper, we have revisited the experimental evidence in support of the phenotypic switching hypothesis using mathematical models for guidance. We showed that the reversible expression of markers after sorting can be explained by assuming that putative CSC markers are not perfect, without invoking phenotypic switching of CCs into CSCs. To illustrate this point we have constructed a hierarchical cancer model in which CSCs can self-renew and give rise to CCs which can duplicate for a finite number of times only.

This basic model can then be modified according to the phenotypic switching hypothesis, introducing a small probability for CCs to transform back into CSCs, or to the imperfect marker hypothesis, introducing probabilities for CSCs and CCs to yield a progeny that is positive or negative to the marker. Finally, the model can also be used to test the effect of sorting errors, when CSCs or CCs are assigned to the wrong category by the instrument. In all these cases the CSC marker appears to be reversibly expressed after sorting. The fraction of positive cells reaches the steady-state value even for negative cells. The model also allows to predict the growth of sorted subpopulations, which could in principle be used to discriminate between various possibilities. We have compared the prediction of the model with experimental results reported in Ref. [4], showing that it is not possible to distinguish between the phenotypic switching and the imperfect marker hypothesis. We also notice that in experiments the effect of imperfect markers is likely to be combined with that of sorting errors. We conclude that experiments suggesting phenotypic switching of CCs into CSC could equally well be interpreted assuming that the putative CSC marker is not perfect. In order to have a more conclusive idea on the behaviour of CSCs, it would interesting to follow their kinetics in vivo in analogy with what is currently done for stem cells[24,25].

## Methods

Recursion equations (5,8,9,10) are solved numerically using a Fortran code. We first reach the steady-state by iterating the equations for at least $N = 100$ steps starting from an initial condition with a single CSC. The steady-state is, however, independent from the initial conditions. Next, we perform the sorting by separating positive and negative cells according to the model and iterate again the equations from the new initial condition. The process is repeated for different parameter values.

1. Fidler, I. J. Tumor heterogeneity and the biology of cancer invasion and metastasis. *Cancer Res* **38**, 2651–60 (1978).

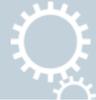

2. Bonnet, D. & Dick, J. E. Human acute myeloid leukemia is organized as a hierarchy that originates from a primitive hematopoietic cell. *Nat Med* **3**, 730–737 (1997).

3. La Porta, C. Cancer stem cells: lessons from melanoma. *Stem Cell Rev* **5**, 61–5 (2009).

4. Gupta, P. B. *et al.* Stochastic state transitions give rise to phenotypic equilibrium in populations of cancer cells. *Cell* **146**, 633–44 (2011).

5. Brock, A., Chang, H. & Huang, S. Non-genetic heterogeneity–a mutation-independent driving force for the somatic evolution of tumours. *Nat Rev Genet* **10**, 336–42 (2009).

6. Turner, B. M. Epigenetic responses to environmental change and their evolutionary implications. *Philos Trans R Soc Lond B Biol Sci* **364**, 3403–18 (2009).

7. Quintana, E. *et al.* Phenotypic heterogeneity among tumorigenic melanoma cells from patients that is reversible and not hierarchically organized. *Cancer Cell* **18**, 510–23 (2010).

8. Roesch, A. *et al.* A temporarily distinct subpopulation of slow-cycling melanoma cells is required for continuous tumor growth. *Cell* **141**, 583–94 (2010).

9. Schatton, T. *et al.* Identification of cells initiating human melanomas. *Nature* **451**, 345–349 (2008).

10. Held, M. A. *et al.* Characterization of melanoma cells capable of propagating tumors from a single cell. *Cancer Res* **70**, 388–97 (2010).

11. La Porta, C. A. M., Zapperi, S. & Sethna, J. P. Senescent cells in growing tumors: population dynamics and cancer stem cells. *PLoS Comput Biol* **8**, e1002316 (2012).

12. Harris, T. E. *The theory of branching processes* (Dover, New York, 1989).

13. Kimmel, M. & Axelrod, D. E. *Branching processes in biology* (Springer, New York, 2002).

14. Vogel, H., Niewisch, H. & Matioli, G. The self renewal probability of hemopoietic stem cells. *J Cell Physiol* **72**, 221–8 (1968).

15. Matioli, G., Niewisch, H. & Vogel, H. Stochastic stem cell renewal. *Rev Eur Etud Clin Biol* **15**, 20–2 (1970).

16. Potten, C. S. & Morris, R. J. Epithelial stem cells in vivo. *J Cell Sci Suppl* **10**, 45–62 (1988).

17. Clayton, E. *et al.* A single type of progenitor cell maintains normal epidermis. *Nature* **446**, 185–9 (2007).

18. Antal, T. & Krapivsky, P. L. Exact solution of a two-type branching process: clone size distribution in cell division kinetics. *Journal of Statistical Mechanics: Theory and Experiment* **2010**, P07028 (2010).

19. Itzkovitz, S., Blat, I. C., Jacks, T., Clevers, H. & van Oudenaarden, A. Optimality in the development of intestinal crypts. *Cell* **148**, 608–19 (2012).

20. Michor, F. *et al.* Dynamics of chronic myeloid leukaemia. *Nature* **435**, 1267–70 (2005).

21. Ashkenazi, R., Gentry, S. N. & Jackson, T. L. Pathways to tumorigenesis–modeling mutation acquisition in stem cells and their progeny. *Neoplasia* **10**, 1170–1182 (2008).

22. Michor, F. Mathematical models of cancer stem cells. *J Clin Oncol* **26**, 2854–61 (2008).

23. Tomasetti, C. & Levy, D. Role of symmetric and asymmetric division of stem cells in developing drug resistance. *Proc Natl Acad Sci U S A* **107**, 16766–71 (2010).

24. Sato, T. *et al.* Paneth cells constitute the niche for lgr5 stem cells in intestinal crypts. *Nature* **469**, 415–8 (2011).

25. Snippert, H. J. *et al.* Intestinal crypt homeostasis results from neutral competition between symmetrically dividing lgr5 stem cells. *Cell* **143**, 134–44 (2010).

26. Lopez-Garcia, C., Klein, A. M., Simons, B. D. & Winton, D. J. Intestinal stem cell replacement follows a pattern of neutral drift. *Science* **330**, 822–5 (2010).

27. Naumov, G. N. *et al.* Cellular expression of green fluorescent protein, coupled with high-resolution in vivo videomicroscopy, to monitor steps in tumor metastasis. *J Cell Sci* **112 (Pt 12)**, 1835–42 (1999).

28. La Porta, C. A. M. Mechanism of drug sensitivity and resistance in melanoma. *Curr Cancer Drug Targets* **9**, 391–397 (2009).

29. Idowu, M. O. *et al.* Cd44+/cd24àì/low cancer stem/progenitor cells are more abundant in triple-negative invasive breast carcinoma phenotype and are associated with poor outcome. *Human Pathology* – (2011).

30. Monzani, E. *et al.* Melanoma contains CD133 and ABCG2 positive cells with enhanced tumourigenic potential. *Eur J Cancer* **43**, 935–946 (2007).

31. Klein, W. M. *et al.* Increased expression of stem cell markers in malignant melanoma. *Mod Pathol* **20**, 102–7 (2007).

32. Hadnagy, A., Gaboury, L., Beaulieu, R & Balicki, D. SP analysis may be used to identify cancer stem cell populations. *Exp Cell Res* **312**, 3701–10 (2006).

33. Keshet, G. I. *et al.* MDR1 expression identifies human melanoma stem cells. *Biochem Biophys Res Commun* **368**, 930–6 (2008).

34. Quintana, E. *et al.* Efficient tumour formation by single human melanoma cells. *Nature* **456**, 593–8 (2008).

35. Boiko, A. D. *et al.* Human melanoma-initiating cells express neural crest nerve growth factor receptor CD271. *Nature* **466**, 133–137 (2010).

36. Taghizadeh, R. *et al.* CXCR6, a newly defined biomarker of tissue-specific stem cell asymmetric self-renewal, identifies more aggressive human melanoma cancer stem cells. *PLoS One* **5**, e15183 (2010).

37. Hoek, K. S. & Goding, C. R. Cancer stem cells versus phenotype-switching in melanoma. *Pigment Cell Melanoma Res* **23**, 746–59 (2010).

## Acknowledgements

We thank J. P. Sethna for useful discussions.

## Author contributions

CAMLP and SZ designed the project and wrote the paper, SZ performed numerical simulations and prepared the figures.

## Additional information

**Competing financial interests:** The authors declare no competing financial interests.

                                                                7